# Filter and nested-lattice code design for fading MIMO channels with side-information

Shih-Chun Lin, Pin-Hsun Lin, Chung-Pi Lee and Hsuan-Jung Su*

**Abstract**

Linear-assignment Gel'fand-Pinsker coding (LA-GPC) is a coding technique for channels with interference known only at the transmitter, where the known interference is treated as side-information (SI). As a special case of LA-GPC, dirty paper coding has been shown to be able to achieve the optimal interference-free rate for interference channels with perfect channel state information at the transmitter (CSIT). In the cases where only the channel distribution information at the transmitter (CDIT) is available, LA-GPC also has good (sometimes optimal) performance in a variety of fast and slow fading SI channels. In this paper, we design the filters in nested-lattice based coding to make it achieve the same rate performance as LA-GPC in multiple-input multiple-output (MIMO) channels. Compared with the random Gaussian codebooks used in previous works, our resultant coding schemes have an algebraic structure and can be implemented in practical systems. A simulation in a slow-fading channel is also provided, and near interference-free error performance is obtained. The proposed coding schemes can serve as the fundamental building blocks to achieve the promised rate performance of MIMO Gaussian broadcast channels with CDIT or perfect CSIT.

## I. INTRODUCTION

Gel'fand and Pinsker [1] first considered the issue of communication with interference *noncausally* available at the transmitter but *not* available at the receiver. Recently, many renewed interests arose in the applications of a subclass of this problem called the linear-assignment Gel'fand-Pinsker coding (LA-GPC), where a linear strategy is used [2]. Costa [2] [3] first applied the LA-GPC in additive Gaussian noise channels, and revealed a surprising result that by treating the Gaussian interference as the side information (SI), the interference-free rate is achievable even when the SI is known only at the transmitter. Costa named this special case of the LA-GPC as dirty paper coding (DPC). The DPC result is based on the assumption that *perfect* channel state information at the transmitter (CSIT) is available. That is, the fading coefficients of the wireless channel are perfectly known not only to the receiver but also to the transmitter. However, it is hard to have perfect CSIT in the wireless setting. Typically, the channel coefficients are estimated at the receiver and fed back with limited feedback channel bandwidth to the transmitter. In practice, we can assume that only the channel distribution information at transmitter (CDIT) is known and adopt the general LA-GPC. For scalar slow fading SI channels, LA-GPC was shown to have the *interference-free outage performance* [4]. For fast fading channels, the LA-GPC also has good (sometimes near optimal) rate performance in scalar and multiple-input multiple-output (MIMO) settings [4] [5].

One of the most important applications of the LA-GPC is the MIMO Gaussian broadcast channel (GBC). A MIMO GBC system consists of one transmitter sending information to many receivers, all equipped with multiple antennas. With perfect CSIT, the *capacity region* of MIMO GBC was shown to coincide with the achievable rate region when MIMO DPC is utilized [6]–[10]. The key to this capacity-achieving performance is that MIMO DPC can efficiently use the information of the multi-user interference, known at the transmitter, to make the receiver decode messages with a rate as if the undesirable interference does not exist. With only CDIT, the MIMO DPC does not perform well [11]. Using the general MIMO LA-GPC has been shown to have significant rate gains over applying the MIMO DPC naively and other beamforming-based strategies in the ergodic fast fading MIMO GBC [11] [5]. In the scalar slow fading GBC, using LA-GPC also provides a significant gain over the time-division scheme [4]. In contrast to DPC for which structured codebook designs are well known [8]–[10], [12], all current promising results of the LA-GPC [4], [5], [11] are based on *unstructured* random Gaussian codebooks. Lack of *structured* codebooks so far hinders practical applications of the LA-GPC.

In this paper, we show that with judiciously designed spatial filters, good nested-lattice coding can achieve all achievable rates of the MIMO LA-GPC. Unlike codebooks used in the previous works [1]–[5], [11], lattice codebook has an *algebraic structure* and is possible to be implemented in practice. We rewrite the LA-GPC rate function in a non-trivial equivalent form. This new form motivates the subtle selections of the transmitter SI filter and the receiver filter to achieve the LA-GPC rate with lattice codes. We also provide a simulation for slow-fading channels, and *near optimal interference-free* error performance is obtained. Our coding can be directly applied to fading MIMO GBC with CDIT to obtain the rate gains derived in [4] [5] [11]. As a by product, we also propose a new structured MIMO DPC which achieves the *optimal* interference-free rate when perfect CSIT is available. Thus while being applied to the MIMO GBC systems with perfect CSIT [7], our MIMO DPC is superior to other existing sub-optimal works [13]–[15]. In summary, the main contributions of this work are

1) We provide the methodology to construct the SI and receiver filters in the nested-lattice coding to make it achieve the LA-GPC rate. This rate was achievable previously only with *unstructured* random Gaussian codebooks [1]–[5], [11]. Although in [1]–[5], [11] LA-GPC and DPC seemed only different in their "linear-assignment matrix" selections in the strategy function, this difference will in fact change the entire random codebook design and the decoding rule [11]. In other words, naively using DPC (designed with perfect CSIT) and dealing with the fading statistics separately for the fading SI channels with CDIT is not a good approach, and will result in a rate strictly lower than the one achieved by the LA-GPC. An example of this rate loss can be found in [11, Section IV.B], and more discussions will be given in Section III and VI. In this work, we show that with lattice coding, the receiver filter must be different from the transmitter SI filter for fading channels. Such result contrasts with the common practice in the lattice-based DPC [8], and also verifies the above observations. Our derivations are new even in the scalar case, and our numerical examples validate this result. These numerical examples are the first real implementations having near optimal performance with *finite codelengths* in SI channels. All prior simulation results [8]–[10], [12] with such performances needed very long codelengths.

2) Our transmitter is subject to a covariance matrix constraint, which is more general than the conventional power constraint over all antennas. An additional transmitter filter is introduced to deal with this new constraint. However, it also incurs new difficulties in the proof of our main result (Theorem 1). The details and comparisons with [16] can be found in the proof. According to our Lemma 2, this filter will make the covariance matrix of the transmitted signal exactly as desired. This result extends the application of the proposed coding to MIMO GBC with general input covariance matrix constraints which subsume the per transmit antenna power constraints [7].

3) As a special case, a new MIMO DPC is also proposed. Since only the filters are adjusted in our

design for different fading conditions, our construction is the first unified design using structured codebooks for MIMO SI channels with perfect CSIT or only CDIT. Currently all other existing MIMO DPC designs, for example, the superposition coding vector DPC [10] or combining scalar DPC with vector channel diagonalization [8], [12], need full CSIT. With only CDIT, these designs all have difficulties to achieve the LA-GPC rate. Our MIMO DPC is also a non-trivial MIMO extension of the scalar one [8], and the detailed comparisons can be found in Section V.

The paper organization is as follows. We define the system model and provide backgrounds on lattice coding in Section II. Section III shows our new form of the LA-GPC rate in Lemma 1. Our main contribution is presented as Theorem 1 in Section IV. Before that, our transmitter filter selection is shown in Lemma 2, while the SI and receiver filter selections are shown in Lemma 3. The detailed comparisons with [8] and the applications of the proposed scheme to MIMO GBC are provided in Section V. Section VI provides some numerical simulation examples. Finally, Section VII concludes this paper.

## II. SYSTEM MODEL AND PRELIMINARIES

### A. *MIMO fading channel with side information at the transmitter*

We focus on the following MIMO channel *

$$\mathbf{y}_t = H_t(\mathbf{x}_t + \mathbf{s}_t) + \mathbf{z}_t, \tag{1}$$

where $t$ is the time index and $1 \leq t \leq T$, $T$ is the number of symbols in the code block; $\mathbf{y}_t \in \mathbb{R}^{N \times 1}$ is the $t$th received symbol, $\mathbf{x}_t \in \mathbb{R}^{M \times 1}$ is the $t$th transmitted vector symbol, $\mathbf{s}_t \in \mathbb{R}^{M \times 1}$ is the $t$th vector interference signal known at the transmitter as the SI, $M$ and $N$ are the number of transmitting and receiving antennas respectively; $H_t \in \mathbb{R}^{N \times M}$ is the random MIMO channel matrix encountered by the transmitted signal to the receiver at time $t$. $\mathbf{z}_t \in \mathbb{R}^{N \times 1}$ is the additive Gaussian noise at the receiver where $\mathbf{z}_t \sim N_{\mathbb{R}}(0, \frac{1}{2}\mathbf{I}_N)$. The channel input is limited by a given input covariance matrix constraint $\frac{1}{2}\Sigma_I$, which is positive semi-definite. This real model can be easily modified to encompass the complex signal model, as shown in Section V.

---

*In this paper, entropy and mutual information are denoted by $h(\cdot)$ and $I(;)$, respectively. Deterministic and random matrices are denoted in bold-face and italic capitals, respectively. For matrix $\mathbf{G}$, $\text{Tr}(\mathbf{G})$ and $\text{Rk}(\mathbf{G})$ denote the trace and rank; $\mathbf{G}^T$ and $\mathbf{G}^\dagger$ denote the transpose and conjugate transpose, respectively. $\mathbf{G}_s^{-1}$ and $|\mathbf{G}_s|$ are the inverse and determinant of a square matrix $\mathbf{G}_s$. And $\mathbf{I}_n$ denotes the identity matrix of dimension $n$. The partial ordering between symmetric matrices are denoted by $\succ$ and $\succeq$, for example, $\mathbf{G}_1 \succeq \mathbf{G}_2$ means $(\mathbf{G}_1 - \mathbf{G}_2)$ is a positive semi-definite matrix. And for a bounded Jordan-measurable region $R \subset \mathbb{R}^m$, $||R||$ denotes the volume of $R$.

With only CDIT, there are two kinds of fading channels considered, the slow and fast fading channels. In the slow fading channels [4], $H_t$ is random but fixed within the codeword length $T$; while in the fast fading channels [4], [5], [11], $H_t, t = 1 \ldots T$, is assumed to be an i.i.d. random process with respect to time. In both cases, $H_t$ can be obtained perfectly at the receiver but only the distribution information is known at the transmitter. We limit the distribution of $\mathbf{s}_t$ to be Gaussian in the channels with only CDIT. For the channels with perfect CSIT [2], [8], $H_t, t = 1 \ldots T$, are constant within the codeword length $T$ and known perfectly at the transmitter. And $\mathbf{s}_t$ can be arbitrarily distributed.

We can rewrite (1) in an equivalent super channel to present our coding scheme more easily in Section IV. By concatenating all $T$ symbols, (1) becomes

$$\mathbf{y} = H(\mathbf{x} + \mathbf{s}) + \mathbf{z}, \qquad (2)$$

where $\mathbf{x} = (\mathbf{x}_1^T, \ldots, \mathbf{x}_T^T)^T$; the noncausally known transmitter SI $\mathbf{s}$ and the noise term $\mathbf{z}$ are obtained similarly from $\mathbf{s}_t$ and $\mathbf{z}_t$ respectively as $\mathbf{x}$ from $\mathbf{x}_t$. The covariance matrix of $\mathbf{z}$ is $\frac{1}{2}\mathbf{I}_{NT}$. The dimension of the real block-diagonal channel matrix $H$ is $NT \times MT$, with its $t$th diagonal term as $H_t$. We also form the channel input covariance matrix constraint as

$$\Sigma_G = \frac{1}{2}\mathbf{I}_T \otimes \Sigma_I, \qquad (3)$$

where $\otimes$ denotes the Kronecker product. It means that the same constraint applies to all vector symbols within a codeword. The input covariance constraint is $\Sigma_G \succeq \Sigma_x$, where $\Sigma_x$ is the covariance matrix of $\mathbf{x}$.

*B. Review of lattices and lattice quantization noise*

An $m_L$-dimensional real lattice $\Lambda$ is defined as $\Lambda = \{\mathbf{Gb} : \mathbf{b} \in \mathbb{Z}^{m_L}\}$, where $\mathbf{G}$ is the $m_L \times m_L$ generator matrix of $\Lambda$. We assume that $\mathbf{G}$ is full rank as in [16], and the lattice is nondegenerate [17]. Let $\Omega$ be any fundamental region [18] of $\Lambda$, the lattice quantizer associated with $\Omega$ with quantizer input $\mathbf{g}$ is defined as $Q_\Omega(\mathbf{g}) = \lambda$, if $\mathbf{g} \in \lambda + \Omega$. The modulo-$\Lambda$ operation associated with $\Omega$ is then

$$\mathbf{g} \bmod_\Omega \Lambda = \mathbf{g} - Q_\Omega(\mathbf{g}). \qquad (4)$$

Let $\mathbf{u}$ be a dither uniformly distributed in $\Omega$ and independent of $\mathbf{g}$. It is proved in [18, Lemma 1] that the dithered quantization error $(\mathbf{g} + \mathbf{u}) \bmod_\Omega \Lambda$ is also uniformly distributed in $\Omega$ as $\mathbf{u}$, and independent of $\mathbf{g}$. The autocorrelation matrix of this error is $\Sigma_\Omega = \mathbb{E}\{\mathbf{u}(\mathbf{u})^T\}$. Since the lattice is nondegenerate, $\Sigma_\Omega$ is positive-definite and nonsingular. One important fundamental region of $\Lambda$ is the Voronoi region $\mathcal{V}$, which is the set of points $\mathbf{g} \in \mathbb{R}^{m_L}$ that are closest to $\mathbf{0}$ in Euclidean distance than to other lattice points $\lambda \in \Lambda$. The second moment [18] associated with this region is denoted as $P(\mathcal{V})$.

## III. MIMO LINEAR-ASSIGNMENT GEL'FAND-PINSKER CODING AND ITS ACHIEVABLE RATE

In this section, we will introduce the MIMO LA-GPC and its achievable rate, denoted as $R_{LA}$, using *random Gaussian codebooks*. The new formula of $R_{LA}$ in Lemma 1 of subsection III-A will play an important role in building surprising connections between $R_{LA}$ and the achievable rate of the proposed coding in Section IV. To illustrate the MIMO LA-GPC, we presents the following channel as in Fig. 1 (a), which represents (1) in the Shannon random-coding setting as [†].

$$Y^N = \underline{H}(X^M + S^M) + Z^N, \tag{5}$$

where $\underline{H}$ is an $N \times M$ random matrix. For simplicity, we first consider the full CSIT case. Without loss of generality, we can replace $\underline{H}$ with deterministic $\mathbf{H}$ as in Fig. 1 (b). Using binning technique on the random codebook [1], the rate

$$I(U^M; Y^N) - I(U^M; S^M) \tag{6}$$

is achievable for any particular choice of $p(u|s)$ and $f(\cdot)$, where $U^M$ is an auxiliary random vector with distribution specified by the conditional distribution $p(u|s)$, and $f(\cdot)$ is a deterministic strategy function such that $X^M = f(U^M, S^M)$. The LA-GPC uses the following "linear-assignment" strategy with random Gaussian codebooks as

$$U^M = \mathbf{W_B} S^M + X^M, \tag{7}$$

where $X^M \sim N_{\mathbb{R}}(0, \frac{1}{2}\Sigma_I)$ is independent of $S^M$, and $\mathbf{W_B}$ is an $M \times M$ matrix. Note that this strategy specifies the function $X^M = f(U^M, S^M)$ as $X^M = U^M - \mathbf{W_B} S^M$. Costa showed that if $\mathbf{W_B}$ can be selected according to the full CSIT $\mathbf{H}$, then the optimal interference-free rate is achievable [2], [3]. He then named this special LA-GPC as DPC.

We now consider channels with only CDIT as in (5). The ergodic fast fading case is first considered, where the channel random process is assumed to be i.i.d. for every time slot [4], [5], [11]. The optimal strategy for this channel is still an open problem due to lack of full CSIT. Thus [4], [5], [11] focused on the the achievable rate

$$I(U^M;Y^N,\underline{H}) - I(U^M;S^M), \qquad (8)$$

with the "linear-assignment" selection (7). The maximum of (8) over all linear assignment matrix $\mathbf{W_B}$ calculated *with only CDIT* is called the "linear-assignment" capacity. Although only the selection of matrix $\mathbf{W_B}$ is different compared with the DPC, this change will change the random binning codebook design. Moreover, the decoder will also be different. The decoder in [2], [3] seeks a codeword that is jointly typical with $Y^N$, while the LA-GPC decoder in ergodic fast fading channel seeks a codeword that is jointly typical with both $Y^N$ and $\underline{H}$ [11, Sec. IV-B]. In both scalar and MIMO fast fading SI channels, the linear-assignment capacity is close to optimal in some signal-to-noise ratio (SNR) regions [4], [5].

---

[†]To emphasize the differences between the lattice codebook setting and the unstructured Shannon random codebook setting, signal vectors in the former are denoted in bold-face lower-cases while those in the latter are denoted in italic capitals with the superscripts specifying their dimensions.

For the quasi-static slow fading channel [4], the decoding error probability cannot be arbitrarily small since the transmitter does not know the reliable transmission rate with only CDIT. In this channel, the outage probability [19] for a given transmission rate $R$ is a better metric than the Shannon capacity to measure the performance. Define $R_{LA}(\underline{\mathbf{H}}) \triangleq I(U^M;Y^N|\underline{H}=\underline{\mathbf{H}}) - I(U^M;S^M)$, this probability is

$$\mathbb{P}\{\underline{H}: R_{LA}(\underline{H}) < R\}, \qquad (9)$$

where $\underline{\mathbf{H}}$ is a realization of $\underline{H}$. In [4], it is shown that LA-GPC achieves the interference-free outage performance in the scalar channel with properly selected $\mathbf{W_B}$ according to the CDIT.

*A. Achievable rate of the MIMO LA-GPC with random Gaussian codebooks*

We now explicitly compute $R_{LA}$, the LA-GPC achievable rate, using *random Gaussian codebooks*. The linear assignment matrix $\mathbf{W_B}$ is assumed to be determined in advance according to the CDIT as in [4], [5]. Since both the achievements of the linear-assignment capacity (8) in ergodic fast fading channels [4], [5], [11] and the outage probability (9) in slow fading channel [4] are based on the coding achieving $R_{LA}$ in a certain channel realization $\underline{H} = \underline{\mathbf{H}}$ [19], we will first focus on this case. Note that this $\underline{\mathbf{H}}$ is only partially known at the transmitter. Also to simplify the presentations in the following sections, as in Section II, we concatenate $T$ random vector $Y^N$ in (5) as

$$Y^{NT} = \mathbf{H}(X^{MT} + S^{MT}) + Z^{NT}, \qquad (10)$$

where the channel realization $\mathbf{H}$ corresponds to $H$ in (2), $X^{MT} \sim N_{\mathbb{R}}(0, \Sigma_G)$ ($\Sigma_G$ is defined in (3)), $S^{MT} \sim N_{\mathbb{R}}(0, \frac{1}{2}\Sigma_s)$, and $Z^{NT} \sim N_{\mathbb{R}}(0, \frac{1}{2}\mathbf{I}_{NT})$, respectively. The covariance matrix $\frac{1}{2}\Sigma_s$ is block-diagonal.

To show the LA-GPC rate, we rewrite channel (10) as

$$Y^{NT} = \mathbf{H}(\sqrt{2}\Sigma_G^*)\tilde{X}^{mT} + \mathbf{H}S^{MT} + Z^{NT} = \tilde{\mathbf{H}}\tilde{X}^{mT} + \mathbf{H}S^{MT} + Z^{NT}, \quad (11)$$

where $\tilde{\mathbf{H}} \triangleq \mathbf{H}\sqrt{2}\Sigma_G^*$ and $\Sigma_G^*$ is an $MT \times \text{Rk}(\Sigma_G)$ matrix which satisfies

$$\Sigma_G^*(\Sigma_G^*)^{\text{T}} = \Sigma_G. \quad (12)$$

The $mT \times 1$ random vector $\tilde{X}^{mT}$ is distributed as $N_{\mathbb{R}}(0, \frac{1}{2}\mathbf{I}_{mT})$ and independent of $S^{MT}$ and $Z^{NT}$, where $mT = \text{Rk}(\Sigma_G)$. Note that $X^{MT} \sim N_{\mathbb{R}}(0, \Sigma_G)$ is distributed the same as $\sqrt{2}\Sigma_G^*\tilde{X}^{mT}$.

We focus on the achievement of the following LA-GPC rate as

$$R_{LA} = \{I(\tilde{U}^{mT}; Y^{NT}) - I(\tilde{U}^{mT}; S^{MT})\}/T, \text{ with } \tilde{U}^{mT} = \mathbf{W}S^{MT} + \tilde{X}^{mT}. \quad (13)$$

Here $\mathbf{W}$ is an $mT \times MT$ block-diagonal matrix satisfying $\sqrt{2}\Sigma_G^*\mathbf{W} = \mathbf{I}_T \otimes \mathbf{W_B}$. The matrix $\mathbf{W_B}$ is computed in advance according to the CDIT as in [5] [4]. Note that $\Sigma_G^*$ is also block-diagonal. Comparing $\tilde{U}^{mT}$ with $U^M$ in (7), we have $\mathbf{1}_T \otimes U^M = \sqrt{2}\Sigma_G^*\tilde{U}^{mT}$, where $\mathbf{1}_T$ is a $T \times 1$ vector with all elements equal to 1. We have

*Lemma 1:* Let $\Sigma_{E_U}$ be the covariance matrix of the linear minimum mean-square error (LMMSE) estimation error $E_{U,MMSE}^{mT}$ to estimate $\tilde{U}^{mT}$ in (13) from $Y^{NT}$ in (11) with LMMSE estimation filter $\mathbf{W}_{\mathbf{U},\text{MMSE}}$, then the LA-GPC achievable rate using random Gaussian codebooks is

$$R_{LA} = \frac{1}{2T} \log \frac{|\frac{1}{2}\mathbf{I}_{mT}|}{|\Sigma_{E_U}|}. \quad (14)$$

*Proof:* From (13),

$$R_{LA} = (h(\tilde{U}^{mT}|S^{MT}) - h(\tilde{U}^{mT}|Y^{NT}))/T. \quad (15)$$

Due to the linear assignment strategy in (13), the first term becomes

$$h(\tilde{U}^{mT}|S^{MT}) = h(\tilde{X}^{mT}|S^{MT}) = h(\tilde{X}^{mT}), \quad (16)$$

where the second equality comes from the independence between $S^{MT}$ and $\tilde{X}^{mT}$. As for the second term, we use the concept of the backward channel in the LMMSE estimation [20] to express $\tilde{U}^{mT}$ as

$$\tilde{U}^{mT} = \mathbf{W}_{\mathbf{U},\text{MMSE}}Y^{NT} + E_{U,MMSE}^{mT}. \quad (17)$$

Then

$$h(\tilde{U}^{mT}|Y^{NT}) = h(\tilde{U}^{mT} - \mathbf{W}_{\mathbf{U},\text{MMSE}}Y^{NT}|Y^{NT}) = h(E_{U,MMSE}^{mT}|Y^{NT}) = h(E_{U,MMSE}^{mT}), \qquad (18)$$

where the last equality comes from the fact that the LMMSE estimation error $E_{U,MMSE}^{mT}$ is independent of $Y^{NT}$ [20]. Using (16) and (18) in (15) and recall that $\tilde{X}^{mT} \sim N_{\mathbb{R}}(0, \frac{1}{2}\mathbf{I}_{mT})$, we have (14). □

## IV. NESTED LATTICE CODING WITH SPATIAL FILTERING

In this section, we will show that combining the proposed spatial filters and "good" nested-lattice coding, $R_{LA}$ in Lemma 1 is still achievable under the transmission input covariance matrix constraint $\Sigma_G$ *without* using random Gaussian codebooks. As in Section III-A, without loss of generality, we focus on the fading channel (2) with a certain realization $H = \mathbf{H}$ as

$$\mathbf{y} = \mathbf{H}(\mathbf{x} + \mathbf{s}) + \mathbf{z}, \qquad (19)$$

where the SI at the transmitter $\mathbf{s} \sim N_{\mathbb{R}}(0, \frac{1}{2}\Sigma_s)$. We assume that $\Sigma_G$ and the linear-assignment matrix $\mathbf{W}$ in (13) are given by [5] [4] according to the available CDIT. We define the nested lattice codes as

*Definition 1:* Let $\Lambda_c$ be a lattice and $\Lambda_q$ be a sublattice of it, that is, $\Lambda_q \subseteq \Lambda_c$. The codeword set of the nested lattice code is $C_c = \{\Lambda_c \bmod \Lambda_q\} \triangleq \{\Lambda_c \cap \mathcal{V}_q\}$, where $\mathcal{V}_q$ is the Voronoi region of $\Lambda_q$.

We choose the code rate of nested lattice code as $R = \frac{1}{T}\log||\mathcal{V}_q||/||\mathcal{V}_c||$, where $\mathcal{V}_c$ is the Voronoi region of $\Lambda_c$. The dimensions of lattices are $mT$, where $mT$ is defined right after (12). Our encoding/decoding scheme is as follows.

**Transmitter:** The transmitter selects a codeword $\mathbf{c_c} \in C_c$ according to the message index and sends

$$\mathbf{x} = \mathbf{F_t}((\mathbf{c_c} - \mathbf{F_s}\mathbf{s} - \mathbf{u}) \bmod \Lambda_q), \qquad (20)$$

where the dither signal $\mathbf{u}$, uniformly distributed in $\mathcal{V}_q$ and independent of the channel, is known to both the transmitter and receiver. The subscript $\mathcal{V}_q$ for the modulo is omitted for brevity, that is, $\mathbf{g} \bmod \Lambda_q = \mathbf{g} - Q_{\mathcal{V}_q}(\mathbf{g}), \forall \mathbf{g} \in \mathbb{R}^{mT}$. The transmitter filter $\mathbf{F_t}$ and the SI filter $\mathbf{F_s}$ will be determined later in Lemma 2 and 3, respectively.

**Decoder:** After passing **x** through the channel in (19), the decoder performs signal processing on the received signal and gets

$$\hat{\mathbf{y}} = \mathbf{L}(\mathbf{F_r}\mathbf{y} + \mathbf{u}), \tag{21}$$

where the receiver filters $\mathbf{F_r}$ and $\mathbf{L}$ will be determined in Lemma 3 and Theorem 1, respectively. We use the generalized minimum Euclidean distance lattice decoder [16] to decode $\mathbf{c_c}$. First the decoder finds

$$\hat{\mathbf{b}} = \arg \min_{\mathbf{b} \in \mathbb{Z}^{mT}} |\hat{\mathbf{y}} - \mathbf{L}\mathbf{G}_c\mathbf{b}|^2, \tag{22}$$

where $\mathbf{G}_c$ is the generator matrix of the channel coding lattice $\Lambda_c$. And the decoded codeword is $\hat{\mathbf{c}}_c = [\mathbf{G}_c\hat{\mathbf{b}}] \bmod \Lambda_q$.

We will show the selection of filters $\mathbf{F_t}$, $\mathbf{F_s}$, $\mathbf{F_r}$ and $\mathbf{L}$ in the following lemmas. First, let the auto-correlation matrix of the dithered quantization error of $\Lambda_q$ be $\Sigma_\mathcal{V}$. Since the lattice $\Lambda_q$ is nondegenerate, $\Sigma_\mathcal{V} \succ \mathbf{0}$. We can apply the Choletsky factorization [21] to obtain the matrix $\Sigma_\mathcal{V}^*$ satisfying

$$\Sigma_\mathcal{V}^*(\Sigma_\mathcal{V}^*)^\mathrm{T} = \Sigma_\mathcal{V}. \tag{23}$$

The matrix $\Sigma_\mathcal{V}^*$ is lower triangular and nonsingular. And we have

*Lemma 2:* Let $\mathbf{F_t} = \Sigma_G^*(\Sigma_\mathcal{V}^*)^{-1}$, where $\Sigma_G^*$ and $\Sigma_\mathcal{V}^*$ are defined in (12) and (23), then the transmitter covariance matrix $\Sigma_x$ satisfies the covariance constraint $\Sigma_G \succeq \Sigma_x$ since $\Sigma_x = \Sigma_G$.

*Proof:* First note that from (4) and (20), the transmitted signal **x** can also be expressed as

$$\mathbf{x} = \mathbf{F_t}(\mathbf{c_q} + \mathbf{c_c} - \mathbf{F_s}\mathbf{s} - \mathbf{u}) = \mathbf{F_t}\tilde{\mathbf{x}}, \tag{24}$$

where $\mathbf{c_q} = Q_{\Lambda_q}(-\mathbf{c_c} + \mathbf{F_s}\mathbf{s} + \mathbf{u})$ and

$$\tilde{\mathbf{x}} \triangleq \mathbf{c_q} + \mathbf{c_c} - \mathbf{u} - \mathbf{F_s}\mathbf{s}. \tag{25}$$

Indeed, $\tilde{\mathbf{x}}$ is the lattice quantization error, which is independent of the interference **s** and distributed as **u** according to Section II-B. Then **x** is distributed as $\mathbf{F_t}\mathbf{u}$. It is zero mean due to the fact that $\mathcal{V}_q = -\mathcal{V}_q$, thus its autocorrelation matrix equals to its covariance matrix $\Sigma_x = \mathbf{F_t}\Sigma_\mathcal{V}\mathbf{F_t}^\mathrm{T}$. With our selection of $\mathbf{F_t}$, $\Sigma_x = \Sigma_G$ due to (12) and (23), and the constraint is satisfied. Note that according to (3), $\Sigma_G \succeq \mathbf{0}$ since $\Sigma_I \succeq \mathbf{0}$. Thus, the full column rank matrix $\Sigma_G^*$ satisfying (12) always exists [21]. □

Now we provide the selections of the SI and receiver filters $\mathbf{F_s}$ and $\mathbf{F_r}$ to make connection between the lattice coding achievable rate and $R_{LA}$ in Lemma 1. These filters are selected according to the linear assignment matrix $\mathbf{W}$ in (13) and the LMMSE filter $\mathbf{W_{U,MMSE}}$ to estimate the auxiliary random vector $\tilde{U}^{mT}$ in Lemma 1 as

*Lemma 3:* Let the filter $\mathbf{F_r}$ be $\sqrt{2}\Sigma_{\mathcal{V}}^*\mathbf{W_{U,MMSE}}$, $\mathbf{F_s}$ be $\sqrt{2}\Sigma_{\mathcal{V}}^*\mathbf{W}$, respectively, where $\Sigma_{\mathcal{V}}^*$ is defined in (23). Then $\hat{\mathbf{y}}$ in (21) equals to $\mathbf{L}(\mathbf{c'_c} + \mathbf{e})$, where $\mathbf{c'_c} \in \Lambda_q + \mathbf{c_c} \in \Lambda_c$ and

$$\mathbf{e} \triangleq (\mathbf{F_r}\tilde{\mathbf{H}}_\mathbf{F} - \mathbf{I}_{mT})\mathbf{u} + (\mathbf{F_r}\mathbf{H} - \mathbf{F_s})\mathbf{s} + \mathbf{F_r}\mathbf{z}. \tag{26}$$

Here $\tilde{\mathbf{H}}_\mathbf{F} \triangleq \mathbf{H}\mathbf{F_t}$ with $\mathbf{F_t}$ specified in Lemma 2. Moreover,

$$\frac{1}{2T} \log \frac{|\Sigma_{\mathcal{V}}|}{|\Sigma_E|} = R_{LA}, \tag{27}$$

where $\Sigma_E$ is the covariance matrix of $\mathbf{e}$ and $\Sigma_{\mathcal{V}}$ is defined in (23).

*Proof:* The proof of $\hat{\mathbf{y}} = \mathbf{L}(\mathbf{c'_c} + \mathbf{e})$ is shown in Appendix A. As for (27), we first let $\mathbf{F_r} = \sqrt{2}\Sigma_{\mathcal{V}}^*\mathbf{W_r}$ where $\mathbf{W_r}$ is an $mT \times NT$ matrix. It will be shown that the optimal $\mathbf{W_r}$ maximizing the left-hand-side (L.H.S.) of (27) is $\mathbf{W_{U,MMSE}}$. First we show that

$$|\Sigma_E| = 2|\Sigma_{\mathcal{V}}||\Sigma_{U'}|, \tag{28}$$

where $\Sigma_{U'}$ is the covariance matrix of

$$E_U^{mT} \triangleq \mathbf{W_r}Y^{NT} - \tilde{U}^{mT}.$$

The Gaussian vectors $Y^{NT}$ and $\tilde{U}^{mT}$ are defined in (11) and (13), respectively. To see this, from the definitions of $Y^{NT}$ and $\tilde{U}^{mT}$, $E_U^{mT}$ equals to

$$(\mathbf{W_r}\tilde{\mathbf{H}} - \mathbf{I}_{mT})\tilde{X}^{mT} + (\mathbf{W_r}\mathbf{H} - \mathbf{W})S^{MT} + \mathbf{W_r}Z^{NT}, \tag{29}$$

where $\tilde{\mathbf{H}}$ is defined right after (11). We observe that both $E_U^{mT}$ and $\mathbf{e}$ are zero mean. The dither $\mathbf{u}$ is uniformly distributed in the Voronoi region $\mathcal{V}_q$ of $\Lambda_q$, thus the covariance matrix of $\mathbf{u}$ is $\Sigma_{\mathcal{V}}$. Since $\tilde{X}^{mT} \sim N(0, \frac{1}{2}\mathbf{I}_{mT})$ by definition, the covariance matrix of $\mathbf{u}$ is equal to the covariance matrix of $\sqrt{2}\Sigma_{\mathcal{V}}^*\tilde{X}^{mT}$. By definition, $\mathbf{s}$ and $\mathbf{z}$ are of the same distributions as $S^{mT}$ and $Z^{NT}$, respectively. Also $\mathbf{u}$, $\mathbf{s}$ and $\mathbf{z}$ are independent. Using these facts, and comparing the chosen $\mathbf{F_r}, \tilde{\mathbf{H}}_\mathbf{F}$, and $\mathbf{F_s}$ in (26) with $\mathbf{W_r}, \tilde{\mathbf{H}}$ and $\mathbf{W}$ in (29), it is easy to check that $\Sigma_E$ equals to the covariance matrix of $\sqrt{2}\Sigma^*_\mathcal{V} E_U^{mT}$. And (28) is valid due to (23).

Now we have $|\Sigma_\mathcal{V}|/|\Sigma_E| = |\frac{1}{2}\mathbf{I}_{mT}|/|\Sigma_{U'}|$ due to (28). Since $E_U^{mT} = \mathbf{W_r} Y^{NT} - \tilde{U}^{mT}$ is the estimation error of $\tilde{U}^{mT}$ from $Y^{NT}$ via the linear transform $\mathbf{W_r}$, choosing $\mathbf{W_r} = \mathbf{W}_{\mathbf{U},MMSE}$ will minimize $|\Sigma_{U'}|$ according to [22, P.2390]. Thus, choosing $\mathbf{W_r} = \mathbf{W}_{\mathbf{U},MMSE}$, the L.H.S. of (27) will be maximized, and $\Sigma_{U'}$ equals to $\Sigma_{E_U}$ in (14). Then (27) is proved. □

Finally, combining the previously specified filters with a "good" nested lattice, the optimality of resultant encoding/decoding scheme is given by the following Theorem. The detailed definition of the "good" nested lattice is omitted, and can be found in [16], [18].

*Theorem 1:* Let filters $\mathbf{F_t}$, $\mathbf{F_s}$ and $\mathbf{F_r}$ be selected as in Lemma 2 and Lemma 3, respectively, and the second moment $P(\mathcal{V}_q)$ of $\Lambda_q$ be $1/2$. If $\mathbf{L}$ in (21) is chosen as $\mathbf{L} = \Sigma^*_\mathcal{V}(\Sigma^*_E)^{-1}$, in which $\Sigma^*_E(\Sigma^*_E)^{\mathrm{T}} = \Sigma_E$ and $\Sigma^*_\mathcal{V}$ is defined in (23), based on sequences of "good" nested lattices, the coding scheme specified in (20)-(22) is able to achieve the LA-GPC rate $R_{LA}$ when $T \to \infty$.

*Proof:* First note that $\Sigma_{E_U}$ is the covariance matrix of the LMMSE error $E_{U,MMSE}^{mT}$ as in Lemma 1, thus it is always invertible [20]. From the Proof of Lemma 3, $\Sigma_E = 2\Sigma_\mathcal{V}\Sigma_{E_U}$, then $\mathbf{L}$ always exists since $\Sigma_\mathcal{V}$ is also invertible. Basically, we will prove that if

$$R < \frac{1}{2T}\log|\mathbf{L}^{\mathrm{T}}\mathbf{L}| = \frac{1}{2T}\log\frac{|\Sigma_\mathcal{V}|}{|\Sigma_E|} = R_{LA}, \quad (30)$$

the specified filters will make the lattice decoding error approach zero as $T \to \infty$. The final equality of (30) was proved in Lemma 3. This proof is a non-trivial extension of [16, Thereom 5], where channels without transmitter SI was considered. Compared with that proof, we propose new filter selection methods tailored for MIMO SI channels with only CDIT as shown in Lemma 1, 2 and 3. Moreover, in [16], the transmitter is subject to a conventional average power constraint. In our case, the filter $\mathbf{F_t}$ designed for a more stringent transmitter covariance matrix constraint $\Sigma_G$ will make the proof to upper-bounding the decoding error probability more involved. The shaping of the lattice quantization noise and its related properties [17] will play an important role in solving this problem. The technical details can be found in Appendix B. □

## V. DISCUSSIONS

As a special case of Theorem 1, we also propose a MIMO DPC for SI channels with perfect CSIT. The optimal linear-assignment matrix $\mathbf{W}$ is $\mathbf{W}_{MMSE}\mathbf{H}$, where $\mathbf{W}_{MMSE}$ is the LMMSE filter used to estimate $\tilde{X}^{mT}$ in (11) with zero interference $S^{MT} = \underline{0}$. When $T \to \infty$, it can easily be checked that with the selected $\mathbf{W}$, $\mathbf{F_r} = \mathbf{W}_{MMSE}$ and $R_{LA}$ becomes the interference-free rate

$$\frac{1}{2T}\log(|\mathbf{H}\Sigma_G\mathbf{H}^T + \frac{1}{2}\mathbf{I}_{NT}|/|\frac{1}{2}\mathbf{I}_{NT}|). \tag{31}$$

If we treat $\mathbf{Hs}$ as SI, then $\mathbf{F_s} = \mathbf{F_r} = \mathbf{W_{MMSE}}$. There are other features that make this MIMO DPC not a straightforward extension of the scalar one [8] [18].

1) Our transmitter is subject to a covariance matrix constraint $\Sigma_G$, instead of the conventional power constraint in [18]. The transmitter filter $\mathbf{F_t}$ is added for this new constraint. The selection of $\mathbf{F_t}$ in Lemma 2 depends on the lattice quantizer chosen. It is more involved than the extension from scalar to MIMO DPC using Gaussian random codebooks in [3], where one can directly set the covariance matrix of the Gaussian random vector (which generates the Gaussian codebook) to $\Sigma_G$ to meet this new constraint. Also due to this constraint, we select $\mathbf{W}_{MMSE}$ according to an equivalent channel $\tilde{\mathbf{H}}$ defined right after (11) instead of the straight-forward one $\mathbf{H}$.

2) With full CSIT, our key observation (27) in the achievement proof is equal to the information-lossless property of the LMMSE estimation [20] in the interference-free channel. Compared with the simple algebra used in [18, pp. 2296] to compute the achievable rate, this property provides new insights to the achievement of the interference-free rate.

3) We chose a different decoder (22) compared to [18]. This lattice decoder can benefit from practical lattice-decoding algorithms [23], which makes the simulations in Section VI possible. Our proof of Theorem 1 is tailored for this lattice decoder, and is completely different from the proof in [18]. In fact, in the MIMO case the equivalent noise $\mathbf{e}$ in (26) is colored. This makes direct extension of the proof in [18] to the MIMO case tedious and difficult. Our proof avoids this problem.

Finally, we briefly sketch the methods to apply the proposed coding to MIMO GBC with full CSIT. For MIMO GBC with CDIT [4], [5], [11], these methods can also be applied easily. Consider a MIMO

GBC system with *K* users and *M* transmitter antennas. The sum of the coded signals of all users will be sent to all receivers. Without loss of generality, we focus on the coding scheme for a user *j*. The *t*th received complex symbol $\mathbf{y}_{j,t}^c$ for this user who has $N_j$ receiver antennas, can be written as

$$\mathbf{y}_{j,t}^c = \mathbf{H}_j^c (\sum_{k=1}^{K} \mathbf{x}_{k,t}^c) + \mathbf{n}_t^c, \tag{32}$$

where $\mathbf{x}_{k,t}^c \in \mathbb{C}^{M \times 1}, 1 \leq k \leq K$, is the *t*th vector symbol of the message of user *k*, $\mathbf{H}_j^c \in \mathbb{C}^{N_j \times M}$ is the MIMO channel gain matrix, and $\mathbf{n}_t^c \in \mathbb{C}^{N_j \times 1}$ is the additive Gaussian noise at the receiver where $\mathbf{n}_t^c \sim N_{\mathbb{C}}(0, \mathbf{I}_{N_j})$. The optimal coding scheme for MIMO GBC [6] will first specify the MIMO DPC achievable rates $R_k$s by determining the encoding order for all users and the covariance matrix constraints $\Sigma_k$s for $\mathbf{x}_{k,t}^c$, and apply the MIMO DPC on each user's message [6], [7]. Whether or how a user's signal will be interfered by the other users' signals is governed by the MIMO DPC encoding order. In general, the signals encoded earlier will be invisible to the signals encoded later, while the former will be interfered by the latter.

Assuming that user *j* is encoded after all the users with indices larger than *j*, it must cancel these interferences. To do this, we rewrite (32) as

$$\mathbf{y}_{j,t}^c = \mathbf{H}_j^c \mathbf{x}_{j,t}^c + \mathbf{H}_j^c (\sum_{k=j+1}^{K} \mathbf{x}_{k,t}^c) + (\mathbf{n}_t^c + \mathbf{H}_j^c (\sum_{k=1}^{j-1} \mathbf{x}_{k,t}^c)). \tag{33}$$

By concatenating the real and imaginary parts of the complex vectors for all *T* symbols similarly to Section II, (33) can be recast as an equivalent real channel fitting (19), where $\Sigma_G$ corresponds $\Sigma_j$, the second term in (33) corresponds to the transmitter SI **Hs**, and the third term corresponds to the noise. Although this noise is not white and Gaussian as in (19), the former property can be resolved by the standard whitening filtering approach, while the latter is met since when $T \to \infty$, **u** approaches Gaussian [17], then all users' transmitted signals approach Gaussian according to the Proof of Lemma 2. From (31), the optimal specified rate $R_j = \log(|\mathbf{I}_{N_j} + \sum_{k \leq j} \mathbf{H}_j^c \Sigma_k (\mathbf{H}_j^c)^\dagger| / |\mathbf{I}_{N_j} + \sum_{k < j} \mathbf{H}_j^c \Sigma_k (\mathbf{H}_j^c)^\dagger|)$ is achievable when $T \to \infty$. Another requirement in [7] is that all users' signals are mutually independent. This requirement is met since for each user, the dither **u** also makes the transmitted signal **x** independent of the interference **s** according to Section II-B. The details of the above statements can be found in [24].

## VI. Numerical examples

In this section, numerical examples are presented to demonstrate the performances of the proposed filters with practical lattice coding schemes. To achieve the rate performance specified in the previous theorems, good nested-lattice encoding/decoding algorithms tailored for a very long codeword length (i.e. $T \to \infty$) are needed. These results may be approached practically by combining the proposed filters with the complex code design methods proposed in [9], [10], [12], which are beyond the scope of this paper. In this section, we alternatively examine the error performance at high SNR with a *reasonable codeword length* (and decoding latency). A Fano sequential-decoding based lattice decoder [25] is used to solve (22) with a good performance.

For simplicity, we consider complex scalar slow fading channels with only CDIT as examples. Using the methods described in Section V, the channel can be recast as a real MIMO 2 by 2 channel. The optimal linear-assignment matrix $\mathbf{W}$ in Section III-A is

$$\mathbf{W} = \frac{1}{\sqrt{2}} \mathbf{I}_T \otimes \begin{bmatrix} 1 - 2^{-R} & 0 \\ 0 & 1 - 2^{-R} \end{bmatrix},$$

where $R$ is the code rate, and the LA-GPC can achieve the interference-free outage performance [4]. The optimal $\mathbf{W}$ for general MIMO slow fading channels is unknown and finding it is very hard and beyond the scope of this paper. The results in [4] were reached using a Gaussian random codebook ensemble with $T \to \infty$. As shown by the simulation results in Fig. 2, with the proposed filters, the interference-free error performance can almost be achieved at high SNR using finite length random lattice codes and decoders in [25]. The fading coefficients are generated as i.i.d. circularly symmetric complex Gaussian random variables with variance equal to 1. As for the lattice ensemble, as in [16], we use the pair of self-similar nested lattices drawn from the ensemble of Construction-A lattices defined in [26]. The parameters of the linear code [26] in this lattice are $(n = 2T, p, k) = (12, 47, 6)$. The lattice codeword length is $T = 6$. A large Gaussian distributed interference signal is added to make SNR 10 dB much larger than the signal-to-interference and noise ratio (SINR). Two different rates, 2 and 4 bits per channel use, are simulated, and the block error rates are obtained by averaging over at least 10000 channel realizations at high SNR. The small gaps between the error curves of random lattice codes and the interference-free outage probabilities in Fig. 2 demonstrate that the decoder decodes as if the interference is almost completely cancelled. For comparison, we also present the "interference as noise" cases. In these cases, the nested-lattice encoder completely ignores the Gaussian SI $\mathbf{s}$, and the decoder treats the interference plus noise $\mathbf{s}+\mathbf{n}$ as an equivalent Gaussian noise to decode the lattice codewords. In [11], applying LA-GPC was shown to have a significant gain over applying DPC naively, where the latter means that the transmitter assumes the channel is fixed at its expected value $E[\underline{H}]$. Since the channel in our simulations is zero mean, the "naive DPC" curve in [11] corresponds to the "interference as noise" curves in our simulations, which also suffer severely due to lack of perfect CSIT.

## VII. CONCLUSIONS

In this paper, we focused on structured codebook designs for fading MIMO side-information channels where interference is known at the transmitter. We showed that the rate performance of the MIMO LA-GPC using random Gaussian codebooks can be achieved by carefully designed spatial filters combined with nested-lattice coding. With only CDIT, the proposed coding scheme has good, sometimes optimal, rate performance. When full CSIT is available, the proposed coding scheme can achieve the optimal interference-free rate. Our coding can be applied to MIMO GBC with CDIT or perfect CSIT to obtain the promised rate performances.

## APPENDIX

### A. Proof of the equivalent channel in Lemma 3

According to the definition of $\mathbf{H_F}$, we can rewrite the channel (19) using $\tilde{\mathbf{x}}$ in (24) as $\mathbf{y} = \tilde{\mathbf{H}}_{\mathbf{F}}\tilde{\mathbf{x}} + \mathbf{H}\mathbf{s} + \mathbf{z}$. Note that the random coding channel (11) has a one-to-one correspondence to this channel. Then

$$\mathbf{F_r}\mathbf{y} + \mathbf{u} = (\tilde{\mathbf{x}} + \mathbf{F_r}\mathbf{H}\mathbf{s} + \mathbf{u}) + (\mathbf{F_r}\tilde{\mathbf{H}}_{\mathbf{F}} - \mathbf{I}_{mT})\tilde{\mathbf{x}} + \mathbf{F_r}\mathbf{z} \quad (34)$$
$$\stackrel{(a)}{=} \mathbf{c_q} + \mathbf{c_c} + (\mathbf{F_r}\mathbf{H} - \mathbf{F_s})\mathbf{s} + (\mathbf{F_r}\tilde{\mathbf{H}}_{\mathbf{F}} - \mathbf{I}_{mT})\tilde{\mathbf{x}} + \mathbf{F_r}\mathbf{z}$$
$$\stackrel{(b)}{=} \mathbf{c_q} + \mathbf{c_c} + \mathbf{e},$$

where equality (a) is due to (25), and (b) is due to the fact that $\tilde{\mathbf{x}}$ distributed as $\tilde{\mathbf{u}}$ as in the Proof of Lemma 2. Let $\mathbf{c'_c} \triangleq \mathbf{c_q} + \mathbf{c_c}$, then $\mathbf{c'_c} \in \Lambda_q + \mathbf{c_c} \in \Lambda_c$ due to the definition of nested lattice. And this concludes the proof.

### B. Proof of Theorem 1

Before introducing the proof, we first borrow the following useful definition from [17]:

*Definition 2:* The shaping of an $m_L$-dimensional lattice quantizer $Q_\mathcal{V}$ (with lattice point $\lambda \in \Lambda$) by an $m_L \times m_L$ nonsingular matrix $\mathbf{F}$ is the quantizer $Q_{\Omega_s}$, for which

$$Q_{\Omega_s}(\mathbf{g}) = \mathbf{F} \cdot Q_\mathcal{V}(\mathbf{F}^{-1}\mathbf{g}).$$

The shaped quantizer $Q_{\Omega_s}$ is also a lattice quantizer with lattice points $\lambda_s = \mathbf{F}\lambda, \lambda \in \Lambda$ and the fundamental region $\Omega_s = \{\mathbf{g} : \mathbf{F}^{-1}\mathbf{g} \in \mathcal{V}\}$, where $\mathcal{V}$ is the Voronoi region of $\Lambda$.

We first consider the achievable rate for any finite $T$, then let $T \to \infty$ to complete the proof. We use an ensemble of dimension $mT$ "good" nested lattices $\{\Lambda_q \subseteq \Lambda_c\}$ defined in [16]. The autocorrelation matrix of the quantization lattice's dithered quantization noise is $\Sigma_\mathcal{V}$. The fundamental volume $||\mathcal{V}_c||$ of the channel coding lattice $\Lambda_c$ is fixed (constant with $T$). As in [16], we use the ambiguity lattice decoder with decision region $\mathcal{D}_{T,\alpha} \triangleq \{\mathbf{g} \in \mathbb{R}^{mT} : |\mathbf{Lg}|^2 \leq \frac{mT}{2}(1+\alpha), \alpha > 0\}$ to simplify the proof. The error probability of the ambiguity lattice decoder will upper-bound that of the generalized minimum Euclidean distance lattice decoder (22). By taking expectation over the ensemble of random channel coding lattices, the average error probability of this decoder is upper-bounded by [26]

$$\mathbb{E}_{\Lambda_c}[P_e(\mathcal{D}_{T,\alpha}|\Lambda_c)] \leq \mathbb{P}(\mathbf{e} \notin \mathcal{D}_{T,\alpha}) + (1+\beta)\frac{||\mathcal{D}_{T,\alpha}||}{||\mathcal{V}_c||}, \ \beta > 0. \tag{35}$$

Let us now focus on the first term in the upper-bound (35). We rewrite it as

$$\mathbb{P}(\mathbf{e} \notin \mathcal{D}_{T,\alpha}) = \mathbb{P}(|\mathbf{Le}|^2 > \frac{mT}{2}(1+\alpha)), \tag{36}$$

where the distribution of $\mathbf{e}$ is shown in Lemma 3. Since $\mathbf{u}$ in (26) is not exactly Gaussian, we will construct a "noisier" Gaussian noise $\mathbf{e_g}$ compared to the non-Gaussian $\mathbf{Le}$ to upper bound (36) as in [16]. However, our construction is more involved than that in [16]. First, due to the additional transmitter filter $\mathbf{F_t}$ in Lemma 2, in our case the noise $\mathbf{e_g}$ must be constructed with the aids of the shaped lattice quantizers. Second, unlike [16], our $\mathbf{e_g}$ is colored and another noise term $\tilde{\mathbf{e}}_\mathbf{g}$ is defined to use the Chernoff bound in [16] to complete the proof. The details come as follows. First note that $\mathbf{u}$ in (26) has covariance matrix $\Sigma_\mathcal{V}$, we rewrite it as

$$(\sqrt{2}\Sigma_\mathcal{V}^*)\mathbf{u}^w, \tag{37}$$

where $\mathbf{u}^w \triangleq \frac{1}{\sqrt{2}}(\Sigma_\mathcal{V}^*)^{-1}\mathbf{u}$. The new quantization noise $\mathbf{u}^w$ is white with autocorrelation matrix $\frac{1}{2}\mathbf{I}_{mT}$, and from [17] we know that $\mathbf{u}^w$ is uniformly distributed in the region

$$\Omega_w \triangleq \{\mathbf{g}: \sqrt{2}\Sigma_\mathcal{V}^*\mathbf{g} \in \mathcal{V}_q\}. \tag{38}$$

This region is the fundamental region if we shape the lattice quantizer associated with $\Lambda_q$ by $\frac{1}{\sqrt{2}}(\Sigma_\mathcal{V}^*)^{-1}$, as described in Definition 2. Using [18, Lemma 11] and following in the footsteps of [16], for all $\mathbf{g} \in \mathbb{R}^{mT}$ the probability density function (pdf) $p_{\mathbf{u}^w}(\mathbf{g})$ of $\mathbf{u}^w$ satisfies

$$p_{\mathbf{u}^w}(\mathbf{g}) \leq (\mathrm{R}_w^c/\mathrm{R}_w^e)^{mT} \exp(o(mT)) p_{n_1}(\mathbf{g}), \tag{39}$$

where the covering radius $\mathrm{R}_w^c$ is the radius of the smallest sphere centered at the origin that contains $\Omega_w$, and $\mathrm{R}_w^e$ is the radius of a sphere having the same volume as $\Omega_w$. The function $p_{n_1}(\mathbf{g})$ is the pdf of a white Gaussian random vector $\mathbf{n}_1 \sim N_\mathbb{R}(0, \sigma^2 \mathbf{I}_{mT})$ with

$$\sigma^2 = (\mathrm{R}_w^c)^2/mT. \tag{40}$$

Now we construct the "noisier" Gaussian error vector corresponding to $\mathbf{Le}$ as

$$\mathbf{e}_g \triangleq \mathbf{L} \cdot \{(\mathbf{F_r}\tilde{\mathbf{H}}_\mathbf{F} - \mathbf{I}_{mT})(\sqrt{2}\Sigma_\mathcal{V}^*)\mathbf{n}_1 + (\mathbf{F_r}\mathbf{H} - \mathbf{F_s})(\mathbf{s}+\mathbf{n}_2) + \mathbf{F_r}(\mathbf{z}+\mathbf{n}_3)\},$$

where $\mathbf{n}_2 \sim N_\mathbb{R}(0, (\sigma^2 - 1/2)\Sigma_s)$, $\mathbf{n}_3 \sim N_\mathbb{R}(0, (\sigma^2 - 1/2)\mathbf{I}_{NT})$. $\mathbf{z}$, $\mathbf{n}_1$, $\mathbf{n}_2$ and $\mathbf{n}_3$ are independent. Since $\mathbf{u}^w$ has covariance matrix $\frac{1}{2}\mathbf{I}_{mT}$, $\frac{1}{2}mT = \mathbb{E}|\mathbf{u}^w|^2 \leq (\mathrm{R}_w^c)^2$ from [17]. From (40), $1/2 \leq \sigma^2$, thus $\mathbf{n}_2$ and $\mathbf{n}_3$ are well-defined. By using (37) in (26) and according to (39), we indeed replace $\mathbf{u}$ in (26) with $\sqrt{2}\Sigma_\mathcal{V}^*\mathbf{n}_1$, and add additional noise vectors $\mathbf{n}_2$ and $\mathbf{n}_3$ to make $\mathbf{e}_g$ "noisier" than $\mathbf{Le}$. Then we can upper-bound (36) as

$$\mathbb{P}(\mathbf{e} \notin \mathcal{D}_{T,\alpha}) \leq (\mathrm{R}_w^c/\mathrm{R}_w^e)^{mT} \exp(o(mT)) \mathbb{P}(|\mathbf{e}_g|^2 \geq \frac{mT}{2}(1+\alpha)). \tag{41}$$

To further upper-bound (41), first note that $\mathbf{e}_g$ is a colored Gaussian vector with covariance matrix

$$\mathbf{L}(2\sigma^2 \Sigma_E)\mathbf{L}^T = 2\sigma^2 \Sigma_\mathcal{V}. \tag{42}$$

The L.H.S. of (42) results from the facts that the distributions of $\mathbf{n}_1$ and $\mathbf{z}+\mathbf{n}_3$ are both $N_\mathbb{R}(0, \sigma^2 \mathbf{I}_{mT})$, and the distribution of $\mathbf{s}+\mathbf{n}_2$ is $N_\mathbb{R}(0, \sigma^2 \Sigma_s)$. And the equality (42) is valid due to the selection $\mathbf{L} = \Sigma_\mathcal{V}^*(\Sigma_E^*)^{-1}$.

Since $\mathbf{e}_g$ is colored, the Chernoff bound used in [16] can not be directly applied. To resolve this issue, we define a white Gaussian vector $\tilde{\mathbf{e}}_g \triangleq (\Sigma_\mathcal{V}^*)^{-1}\mathbf{e}_g$. Since $\Sigma_\mathcal{V}$ is symmetric, from the Rayleigh-Ritz theorem [21], we know that

$$\lambda_{\min}(\Sigma_\mathcal{V}^{-1})\mathbf{e}_g^T\mathbf{e}_g \leq \mathbf{e}_g^T \Sigma_\mathcal{V}^{-1} \mathbf{e}_g = |\tilde{\mathbf{e}}_g|^2, \tag{43}$$

19where $\lambda_{\min}(\Sigma_{\mathcal{V}}^{-1})$ is the minimum eigenvalue of $\Sigma_{\mathcal{V}}^{-1}$. Since $\Sigma_{\mathcal{V}}$ is positive definite, $1/\lambda_{\min}(\Sigma_{\mathcal{V}}^{-1}) = \lambda_{\max}(\Sigma_{\mathcal{V}}) > 0$. We further denote $\lambda_{\max}(\Sigma_{\mathcal{V}})$ as $\lambda_{\max}$ to simplify the notation. From (43), we know

$$\mathbb{P}(|\mathbf{e}_g|^2 \geq \frac{mT}{2}(1+\alpha)) \leq \mathbb{P}\left(|\tilde{\mathbf{e}}_g|^2 \geq \frac{mT}{2\lambda_{\max}}(1+\alpha)\right), \tag{44}$$

since if $|\mathbf{e}_g|^2 \geq mT(1+\alpha)/2$ then $|\tilde{\mathbf{e}}_g|^2 \geq mT(1+\alpha)/2\lambda_{\max}$. From (42) and the definition of $\tilde{\mathbf{e}}_g$, we know that $\tilde{\mathbf{e}}_g \sim N_{\mathbb{R}}(0, 2\sigma^2 \mathbf{I}_{mT})$ and the Chernoff bound in [16] can be applied to bound (44).

Finally, using (41), (44) and following the bounding technique in [16], for arbitrary $\alpha > 0$, $\varepsilon_2 > 0$ and sufficiently large $T$ we have

$$\mathbb{P}(\mathbf{e} \notin \mathcal{D}_{T,\alpha}) \leq \exp\left(-mT\left(-\log\frac{R_w^u}{R_w^e} + \frac{(\alpha'' - 1 - \log \alpha'')}{2} - \frac{o(mT)}{mT}\right)\right) \leq \varepsilon_2/2, \tag{45}$$

where $\alpha'' = (1+\alpha)/(4\sigma^2 \lambda_{\max})$. The last inequality comes from the following facts. First, since the quantization lattice $\{\Lambda_q\}$ is "good" (defined in [16]) with second moment $1/2$, $\Sigma_{\mathcal{V}} \to \frac{1}{2}\mathbf{I}_{mT}$ when $T \to \infty$ [17]. From (38), $\Omega_w \to \mathcal{V}_q$ and $R_w^u/R_w^e$ approaches the covering efficiency [16] of $\{\Lambda_q^T\}$. Then $\log R_w^u/R_w^e \to 0$ since $\{\Lambda_q^T\}$ is "good" [16]. Note that $\Sigma_{\mathcal{V}} \to \frac{1}{2}\mathbf{I}_{mT}$ then $\lambda_{\max} \to 1/2$. Following [16] we know that $\sigma^2 \to 1/2$ and $\alpha'' - 1 - \log \alpha'' > 0$, for some arbitrary $\alpha > 0$. Thus (45) holds.

As for the second term in the upper-bound (35), it can be proved that if the lattice code rate meets (30), then this term can be upper-bounded by $\varepsilon_1/2$ for an arbitrarily small $\varepsilon_1 > 0$ and sufficiently large $T$. The proof is similar to [16] and the details can be found in [24].

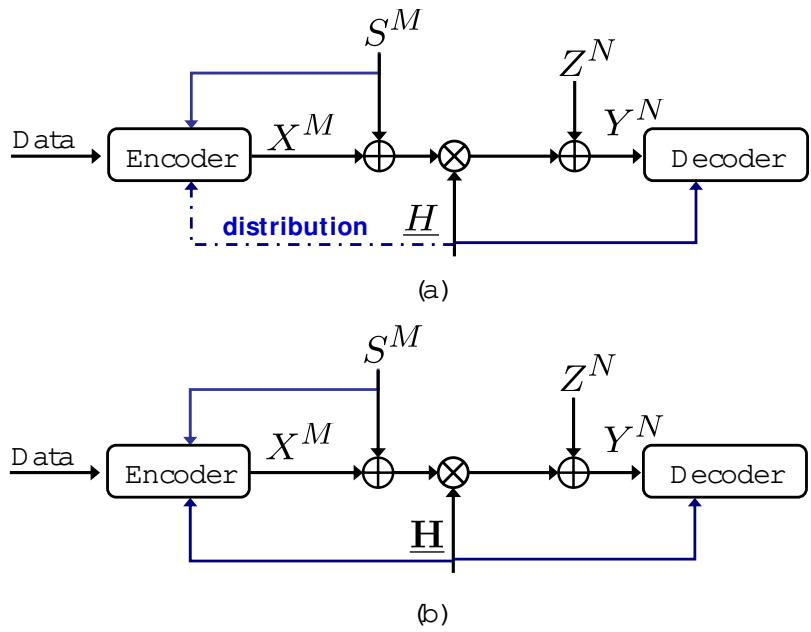

Fig. 1

MIMO side-information channel with (a) CDIT (b) Perfect CSIT.

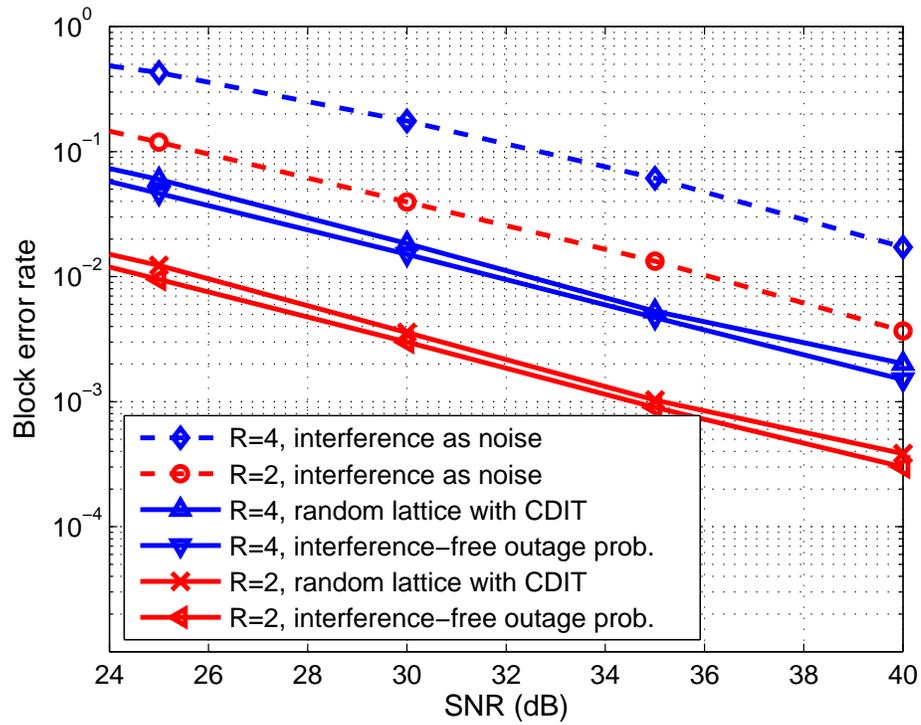

Fig. 2

With the proposed filters, random lattice codes achieve almost interference-free error performance for complex scalar slow fading SI channels with CDIT at high SNR.